\documentclass[12pt]{article}
\usepackage{amsmath}
\newlength{\extraspace}
\newlength{\extraspaces}
\newcommand{\be}{\begin{equation}
\addtolength{\abovedisplayskip}{\extraspaces}
\addtolength{\belowdisplayskip}{\extraspaces}
\addtolength{\abovedisplayshortskip}{\extraspace}
\addtolength{\belowdisplayshortskip}{\extraspace}}
\newcommand{\ee}{\end{equation}}
\newcommand{\ba}{\begin{eqnarray}
\addtolength{\abovedisplayskip}{\extraspaces}
\addtolength{\belowdisplayskip}{\extraspaces}
\addtolength{\abovedisplayshortskip}{\extraspace}
\addtolength{\belowdisplayshortskip}{\extraspace}}
\newcommand{\ea}{\end{eqnarray}}
\newcommand{\nonu}{\nonumber \\[.5mm]}
\newcommand{\A}{&\!\!\!}
\begin{document}

\thispagestyle{empty}
\begin{flushright}
SIT-LP-07/09 \\
{\tt arXiv:0710.1680[hep-th]} \\
September, 2007
\end{flushright}
\vspace{7mm}
\begin{center}
{\large{\bf On vacuum structures of $N = 2$ LSUSY QED \\[2mm] 
equivalent to $N = 2$ NLSUSY model}} \\[7mm]
{\bf (Dedicated to the late Professor Julius Wess)}\\[12mm]
{\sc Kazunari Shima} 
\footnote{
\tt e-mail: shima@sit.ac.jp} \ 
and \ 
{\sc Motomu Tsuda} 
\footnote{
\tt e-mail: tsuda@sit.ac.jp} 
\\[2mm]
{\it Laboratory of Physics, Saitama Institute of Technology \\
Fukaya, Saitama 369-0293, Japan } \\[5mm]
{\sc Wolfdieter Lang}
\footnote{
\tt e-mail: wolfdieter.lang@physik.uni-karlsruhe.de} 
\\[2mm]
{\it Institute for Theoretical Physics, Karlsruhe University, \\
D-76128 Karlsruhe, Germany } \\[12mm]
\begin{abstract}
The vacuum structure of $N = 2$ linear supersymmetry (LSUSY) invariant QED, 
which is equivalent to $N = 2$ nonlinear supersymmetry (NLSUSY) model, 
is studied explicitly in two dimensional space-time ($d = 2$). 
Two different isometries $SO(1,3)$ and $SO(3,1)$ appear for the vacuum field configuration 
corresponding to the various parameter regions. 
Two different field configurations of $SO(3,1)$ isometry describe the two different physical vacua, 
i.e. one breaks spontaneously both $U(1)$ and SUSY and the other breaks spontaneously SUSY alone. \\[3mm]
\end{abstract}
\end{center}

\newpage
\noindent
Supersymmetry (SUSY) in particle field theory \cite{WZ1,VA} is a profound notions 
related to space-time symmetry. 
Therefore, the evidences of SUSY and its spontaneous breakdown \cite{SS,FI,O} should be studied  
not only in (low energy) particle physics but also in cosmology, 
i.e. in a framework necessarily  accomodating graviton. 

In a series of  works along these viewpoints, we have found group theoretically that 
the $SO(10)$ super-Poicar\'e (SP) group may be a unique and minimal group among all $SO(N)$ SP groups, 
which possesses a single irreducible linear (L) SUSY representation accomodating graviton 
and the standard model (SM) with just three generations of quarks and leptons \cite{KS1}. 

The advocated difficulty for constructing non-trivial $N > 8$ SUSY (gravity) theory, 
the so called no-go theorem based on S-matrix argument \cite{CM,HLS} 
can be circumvented by adopting the {\it nonliner (NL) representation} of SUSY \cite{WB}, 
i.e. the vacuum degeneracy of the fundamental action. 
Volkov-Akulov (VA) model \cite{VA} gives the NL representation of SUSY describing the dynamics of spin $1/2$ 
Nambu-Goldstone (NG) fermion accompanying the spontaneous SUSY breaking for $N = 1$.           \par
The fundamental action (called nonlinear supersymmetric general relativity (NLSUSY GR)) 
of empty Einstein-Hilbert (EH) type 
for $N > 8$ SUSY (gravity) theory with $N > 8$ supercharges, 
i.e. the NLSUSY invariant interaction of $N$ NG fermion 
with spin $2$ graviton, has been constructed by extending the geometric arguments 
of Einstein general relativity (EGR) on Riemann space-time to a new space-time just inspired by NLSUSY, 
where tangent space-time is specified  not only by $x_a$ for $SO(1,3)$ 
but also by the Grassmanian $\psi_\alpha$ for isomorphic $SL(2C)$ of NLSUSY \cite{KS2,KS3}. 
The compact isomorphic groups $SU(2)$ and $SO(3)$ for the gauge symmetry 
of 't Hooft-Polyakov monopole are generalized 
to the noncompact isomorphic groups $SO(1,3)$ and $SL(2C)$ for space-time symmetry and the consequent 
NLSUSY GR action possesses promissing large symmetries isomorphic to $SO(10)$ SP \cite{ST3,ST4}.
These results mean that the no-go theorem is overcome (circumvented) in the sense that 
the non-tivial $N$-extended SUSY gravity theory with $N > 8$ has been constructed  
and graviton and $N$ NG fermions with the spin difference $3/2$ can be coupled 
in a SUSY invariant way. 
We think that the geometric arguments of EGR principle 
has been generalized naturally, which accomodates {\it geometrically} spin $1/2$ matter 
as NG fermion accompanying spontaneous SUSY breaking encoded on tangent space-time as NLSUSY. 

NLSUSY GR (called superon-graviton model (SGM) from composite viewpoint) on new empty space-time 
written in the form of the {\it vacuum} EH type 
is unstable due to NLSUSY structure of tangent space-time and  
decays (called {\it Big Decay} \cite{ST4}) spontaneously into ordinary EH action 
with the cosmplogical constant $\Lambda$, NLSUSY action for $N$ NG fermions 
(called {\it superons} as hypothetical spin $1/2$ objects) 
and their gravitational interactions on ordinary Riemann space-time, 
which ignites the Big Bang of the present universe. 
We have shown qualitatively that NLSUSY GR may potentially describe a new paradigm (SGM) 
for the SUSY unification of space-time and matter,  
where particular SUSY compositeness composed of superons for all (observed) particles except the graviton emerges 
as an ultimate feature of nature behind the familiar LSUSY models (MSSM, SUSY GUTs) \cite{KS3,STS} and SM as well. 
That is, all (observed) low energy particles may be eigenstates of $SO(N)$ SP expressed uniquely 
as the SUSY composites of $N$ superons. 

Due to the high nonlinearity of the SGM action we have not yet succeeded in extracting directly 
the evidence of such  (low energy) physical meanings of SGM on curved Riemann space-time.      \\ 
However, considering that SGM action reduces to the $N$-extended NLSUSY action in asymptotic 
Riemann-flat $(e^a{}_\mu \rightarrow \delta^a_\mu)$ space-time after the Big Decay, 
it is interesting from the low energy physics viewpoint to construct the $N$-extended LSUSY theory 
equivalent to the $N$-extended NLSUSY model. 
The relation between $N = 1$ LSUSY representations and $N = 1$ NLSUSY representations 
in flat (Minkowski) space-time is well understood by using the superfield method\cite{WB,IK}. 
The equivalence between $N = 1$ LSUSY ${\it free}$ theory for LSUSY supermultiplet 
and  $N = 1$ NLSUSY VA  model for NG fermion is demonstrated by many authors \cite{R,UZ,STT1} 
and $N = 2$ case as well \cite{STT2}, 
where each field of LSUSY supermultiplet is expressed uniqely as the composite of NG fermions of NLSUSY 
called SUSY invariant relations. 
Consequently we are tempted to imagine some composite structure (far) behind the SM 
and the familiar LSUSY models, e.g. MSSM and SUSY GUT.   \\  
Recently, we have shown explicitly by the heuristic arguments for simplicity 
in two space-time dimensions ($d = 2$) \cite{ST2,ST5} 
that $N = 2$ LSUSY interacting QED is equivalent to $N = 2$ NLSUSY model. 
(Note that the minimal realistic SUSY QED in SGM composite scenario is described by $N = 2$ SUSY \cite{STT2}.) 

In this paper we study explicitly the vacuum structure of $N = 2$ LSUSY QED in the SGM scenario in $d = 2$ \cite{ST5}. 

The $N = 2$ NLSUSY action for two superons (NG fermions) $\psi^i\ (i=1,2)$ in $d = 2$ is written as follows, 
\ba
S_{N=2{\rm NLSUSY}} 
= \A \A -{1 \over {2 \kappa^2}} \int d^2 x \ \vert w \vert
\nonu
= \A \A - {1 \over {2 \kappa^2}} \int d^2 x 
\left\{ 1 + t^a{}_a + {1 \over 2!}(t^a{}_a t^b{}_b - t^a{}_b t^b{}_a) 
\right\} 
\nonu
= \A \A - {1 \over {2 \kappa^2}} \int d^2 x 
\left\{ 1 - i \kappa^2 \bar\psi^i \!\!\not\!\partial \psi^i 
- {1 \over 2} \kappa^4 
( \bar\psi^i \!\!\not\!\partial \psi^i \bar\psi^j \!\!\not\!\partial \psi^j 
- \bar\psi^i \gamma^a \partial_b \psi^i \bar\psi^j \gamma^b \partial_a \psi^j ) 
\right\} 
\nonu
= \A \A - {1 \over {2 \kappa^2}} \int d^2 x 
\left\{ 1 - i \kappa^2 \bar\psi^i \!\!\not\!\partial \psi^i 
\right. 
\nonu
\A \A 
\left. 
- {1 \over 2} \kappa^4 \epsilon^{ab} 
( \bar\psi^i \psi^j \partial_a \bar\psi^i \gamma_5 \partial_b \psi^j 
+ \bar\psi^i \gamma_5 \psi^j \partial_a \bar\psi^i \partial_b \psi^j ) 
\right\}, 
\label{VAaction2}
\ea
where $\kappa^{2}=({c^{4}\Lambda \over 8\pi G})^{-1}$ in the SGM scenario and 
\begin{equation}
\vert w \vert = \det(w^a{}_b) = \det(\delta^a_b + t^a{}_b), 
\ \ \ 
t^a{}_b = - i \kappa^2 \bar\psi^i \gamma^a \partial_b \psi^i. 
\end{equation}

While, the helicity states contained in $d = 2$ $N = 2$ LSUSY QED are 
the vector supermultiplet containing $U(1)$ gauge field  
%
\[\left(\begin{array}{c}
      +{1} \\
\begin{array}{cc}
 +{1 \over 2}, \  +{1 \over 2} 
\end{array} \\
0
\end{array}  \right) + [{\rm CPT\ conjugate}], \]
%
%
and the scalaer supermultiplet for matter fields  
\[\left(\begin{array}{c}
      +{1 \over 2} \\
\begin{array}{cc}
 0 ,\  0 
\end{array} \\
-{1 \over 2}
\end{array}  \right) + [{\rm CPT\ conjugate}]. \]
%
%
The $N = 2$ LSUSY QED action in $d = 2$  for the massless case
is written as follows\cite{ST5}, 
\ba
S_{N=2{\rm SUSYQED}} 
=
\A \A \int d^2 x \left[ - {1 \over 4} (F_{ab})^2 
+ {i \over 2} \bar\lambda^i \!\!\not\!\partial \lambda^i 
+ {1 \over 2} (\partial_a A)^2 
+ {1 \over 2} (\partial_a \phi)^2 
+ {1 \over 2} D^2 
- {1 \over \kappa} \xi D 
\right. 
\nonu 
\A \A
+ {i \over 2} \bar\chi \!\!\not\!\partial \chi 
+ {1 \over 2} (\partial_a B^i)^2 
+ {i \over 2} \bar\nu \!\!\not\!\partial \nu 
+ {1 \over 2} (F^i)^2 
\nonu 
\A \A
+ f ( A \bar\lambda^i \lambda^i + \epsilon^{ij} \phi \bar\lambda^i \gamma_5 \lambda^j 
- A^2 D + \phi^2 D + \epsilon^{ab} A \phi F_{ab} ) 
\nonu 
\A \A
+ e \left\{ i v_a \bar\chi \gamma^a \nu 
- \epsilon^{ij} v^a B^i \partial_a B^j 
+ \bar\lambda^i \chi B^i 
+ \epsilon^{ij} \bar\lambda^i \nu B^j 
- {1 \over 2} D (B^i)^2 \right. 
\nonu
\A \A 
\left. \left. 
+ {1 \over 2} (\bar\chi \chi + \bar\nu \nu) A 
- \bar\chi \gamma_5 \nu \phi \right\}
+ {1 \over 2} e^2 (v_a{}^2 - A^2 - \phi^2) (B^i)^2 \right], 
\label{L2action}
\ea
where  $(v^a, \lambda^i, A, \phi, D)$ ($F_{ab} = \partial_a v_b - \partial_b v_a$) 
is the off-shell vector supermultiplet containing $v^a$ for a $U(1)$ vector field, 
$\lambda^i$ for doublet (Majorana) fermions 
and $A$ for a scalar field in addition to $\phi$ for another scalar field 
and $D$ for an auxiliary scalar field, 
while ($\chi$, $B^i$, $\nu$, $F^i$) is off-shell scalar supermultiplet containing 
$(\chi, \nu)$ for two (Majorana) fermions, 
$B^i$ for doublet scalar fields and $F^i$ for auxiliary scalar fields. 
Also $\xi$ is an arbitrary demensionless parameter giving a magnitude of SUSY breaking mass, 
and $f$ and $e$ are Yukawa and gauge coupling constants with the dimension (mass)$^1$, respectively. 

The equivalence of LSUSY QED action (\ref{L2action}) to NLSUSY action (\ref{VAaction2}) 
up to surface terms was shown explicitly by substituting the (generalized) SUSY invariant relations 
for the component fields of supermultiplet into the LSUSY QED action (\ref{L2action}) \cite{ST5}. 
%
%
$N = 2$ LSUSY QED action (\ref{L2action}) can be rewritten as the familiar manifestly covariant form 
in terms of the complex quantities defined by 
\begin{equation}
\chi_D = {1 \over \sqrt{2}} (\chi + i \nu), 
\ \ \ B = {1 \over \sqrt{2}} (B^1 + i B^2), 
\ \ \ F = {1 \over \sqrt{2}} (F^1 - i F^2).  
\label{cfields}
\end{equation}
The resulting action is manifestly invariant under the  local $U(1)$ transformation 
\ba
\A \A 
(\chi_D, B, F) \ \ \rightarrow \ \ (\chi'_D, B', F')(x) = e^{i \Omega(x)} (\chi_D, B, F)(x), 
\nonu
\A \A 
v_a \ \ \rightarrow \ \ v'_a(x)  = v_a(x) + {1 \over e} \partial_a \Omega(x). 
\label{u1}
\ea
(For further details see ref.\cite{ST5}.)  

For extracting the low energy particle physics contents of $N = 2$ SGM (NLSUSY GR) 
we consider in Riemann-flat asymptotic space-time, where $N = 2$ SGM reduces to 
essentially $N = 2$ NLSUSY action  equivalent to $N = 2$ SUSY QED action, i.e. 
%
\begin{equation}
L_{N=2{\rm SGM}} \overset{e^a{}_\mu \rightarrow \delta^a_\mu}{\longrightarrow} 
L_{N=2{\rm NLSUSY}} + [{\rm suface\ terms}]
= L_{N=2{\rm SUSYQED}}. 
\end{equation}
%
Now we study the vacuum structure of $N = 2$ SUSY QED action (\ref{L2action}). 
The vacuum is determined by the minimum of the potential $V(A, \phi, B^i, D)$, 
\footnote{
The terms, ${1 \over 2} e^2 (A^2 + \phi^2) (B^i)^2$, 
should be added to Eqs.(7) and (8) 
but they do not change the final results. 
}
%
\begin{equation}
V(A, \phi, B^i, D)  =  - {1 \over 2} D^2 + \left\{ {\xi \over \kappa} 
+ f(A^2 - \phi^2) + {1 \over 2} e (B^i)^2 \right\} D.  
\label{potential-1}
\end{equation}
%
Substituting the solution of the equation of motion for the auxiliary field $D$
we obtain 
%
\begin{equation}
V(A, \phi, B^i) = {1 \over 2} f^2 \left\{ A^2 - \phi^2 + {e \over 2f} (B^i)^2 
+ {\xi \over {f \kappa}} \right\}^2 \ge 0. 
\label{potential-2}
\end{equation}
%

The configurations of the fields corresponding to the vacua in $(A, \phi, B^i)$-space, 
which are $SO(1,3)$ or $SO(3,1)$ invariant,   
are classified according to the signatures of the parameters $e, f, \xi, \kappa$ as follows: \\[2mm]
(I) For $-ef > 0$, \ \ $-{\xi \over {f \kappa}} > 0$ case, 
%
\begin{equation}
A^2 - \phi^2 - (\tilde B^i)^2 = k^2. 
\ \ \ \left( \tilde B^i = \sqrt{-{e \over 2f}} B^i, \ \ 
k^2 = -{\xi \over {f \kappa}} \right) 
\end{equation}
%
(II) For $-ef < 0$, \ \ $-{\xi \over {f \kappa}} > 0$ case, 
%
\begin{equation}
A^2 - \phi^2 + (\tilde B^i)^2 = k^2. 
\ \ \ \left( \tilde B^i = \sqrt{e \over 2f} B^i, \ \ 
k^2 = -{\xi \over {f \kappa}} \right) 
\end{equation}
%
(III) For $-ef > 0$, \ \ $-{\xi \over {f \kappa}} < 0$ case, 
%
\begin{equation}
- A^2 + \phi^2 + (\tilde B^i)^2 = k^2. 
\ \ \ \left( \tilde B^i = \sqrt{-{e \over 2f}} B^i, \ \ 
k^2 = {\xi \over {f \kappa}} \right) 
\end{equation}
%
(IV) For $-ef < 0$, \ \ $-{\xi \over {f \kappa}}< 0$ case, 
%
\begin{equation}
- A^2 + \phi^2 - (\tilde B^i)^2 = k^2. 
\ \ \ \left( \tilde B^i = \sqrt{e \over 2f} B^i, \ \ 
k^2 = {\xi \over {f \kappa}} \right) 
\end{equation}
%
We find that the vacua (I) and (IV) with $SO(1,3)$ isometry in $(A, \phi, B^i)$-space are unphysical, 
for they produce pathological wrong sign kinetic terms for the fields expanded around the vacuum. 

As for the cases (II) and (III) we perform similar arguments as shown below 
and find that two different physical vacua appear. 
%
%
%
The physical particle spectrum is obtained by expanding the field $(A, \phi, B^i)$ around the vacuum 
with $SO(3,1)$ isometry.              \par
For case (II), the following two expressions (IIa) and (IIb) are considered: \\
Case (IIa) 
\[
\begin{array}{lll}
A \A = (k + \rho)\sin\theta \cosh\omega, \A {}      
\\
\phi \A = (k + \rho) \sinh\omega, \A {}
\\
\tilde B^1 \A = (k + \rho) \cos\theta \cos\varphi \cosh\omega, \A{}
\\
\tilde B^2 \A = (k + \rho) \cos\theta \sin\varphi \cosh\omega \A {}
\end{array}
\]
and \\
Case (IIb) 
\[
\begin{array}{lll}
A &\!\!\! = - (k + \rho) \cos\theta \cos\varphi \cosh\omega, &\!\!\! {}
\\
\phi &\!\!\! = (k + \rho) \sinh\omega, &\!\!\! {}
\\
\tilde B^1 &\!\!\! = (k + \rho) \sin\theta \cosh\omega, &\!\!\ {}
\\
\tilde B^2 &\!\!\! = (k + \rho) \cos\theta \sin\varphi \cosh\omega. &\!\!\! {}
\end{array}
\]
%
%
%
%
%
Note that for the case (III) the arguments are the same by exchanging $A$ and $\phi$, 
which we call (IIIa) and (IIIb). 

Substituting these expressions into $ L_{N=2{\rm SUSYQED}}(A, \phi, B^i)$ and 
expanding the action around the vacuum configuration  we obtain the physical particle contents. 
For the cases (IIa) and (IIIa) we obtain 
\ba
L_{N=2{\rm SUSYQED}} 
\A =\A {1 \over 2} \{ (\partial_a \rho)^2 - 2 ef k^2 \rho^2 \} 
\nonu
\A \A 
+ {1 \over 2} \{ (\partial_a \theta)^2 + (\partial_a \omega)^2 - 2 ef k^2 (\theta^2 + \omega^2) \} 
\nonu
\A \A 
+ {1 \over 2} (\partial_a \varphi)^2 
\nonu
\A \A 
- {1 \over 4} (F_{ab})^2 + ef k^2 v_a^2 
\nonu
\A \A 
+ {i \over 2} \bar\lambda^i \!\!\not\!\partial \lambda^i 
+ {i \over 2} \bar\chi \!\!\not\!\partial \chi 
+ {i \over 2} \bar\nu \!\!\not\!\partial \nu 
+ \sqrt{2ef} (\bar\lambda^1 \chi - \bar\lambda^2 \nu) 
+ \cdots, 
\ea
and the consequent mass genaration 
\ba
\A \A 
m_\rho^2 = m_\theta^2 = m_\omega^2 = m_{v_a}^2 = 2 ef k^2  
= - {{2 \xi e} \over \kappa}, 
\nonu
\A \A 
m_{\lambda^{i}} = m_{\chi} =  m_{\nu}= m_{\varphi}= 0. 
\nonu
\A \A 
%
\ea
(Note that $\varphi$ is the NG boson for the spontaneous breaking of $U(1)$ symmetry, i.e. the $U(1)$ phase of $B$,  
and totally gauged away by the Higgs-Kibble mechanism with $\Omega(x) = \sqrt{ e \kappa/2}\varphi(x)$ 
for the $U(1)$ gauge (\ref{u1}).) 
The vacuum breaks both SUSY and the local $U(1)$ spontaneously. 
All bosons have the same mass which is different from the cases (IIb) and (IIIb) 
and remarkably all fermions remain massless. 
The physical origin of the off-diagonal mass terms 
$\sqrt{2ef} (\bar\lambda^1 \chi - \bar\lambda^2 \nu) 
= \sqrt{2ef} (\bar\chi_{\rm D} \lambda + \bar\lambda \chi_{\rm D})$ 
($\lambda \sim \lambda^1 - i \lambda^2$) 
for fermions is unclear, which would induce mixings of fermions 
and/or the lepton (baryon) number violations.  
%
%

By similar computations for (IIb) and (IIIb) we obtain 
\ba
L_{N=2{\rm SUSYQED}} 
\A = \A {1 \over 2} \{ (\partial_a \rho)^2 - 4 f^2 k^2 \rho^2 \} 
\nonu
\A \A 
+ {1 \over 2} \{ (\partial_a \theta)^2 + (\partial_a \varphi)^2 
- e^2 k^2 (\theta^2 + \varphi^2) \} 
\nonu
\A \A 
+ {1 \over 2} (\partial_a \omega)^2 
\nonu
\A \A 
- {1 \over 4} (F_{ab})^2 
\nonu
\A \A 
+ {1 \over 2} (i \bar\lambda^i \!\!\not\!\partial \lambda^i 
- 2 f k \bar\lambda^i \lambda^i) 
\nonu
\A \A 
+ {1 \over 2} \{ i (\bar\chi \!\!\not\!\partial \chi + \bar\nu \!\!\not\!\partial \nu) 
- e k (\bar\chi \chi + \bar\nu \nu) \} 
+ \cdots. 
\ea
%
%
and the following mass spectrum which indicates that SUSY is broken spontaneously as expected; 
\ba
\A \A 
m_\rho^2 = m_{\lambda^i}^2 = 4 f^2 k^2 = -{{4 \xi f} \over \kappa}, 
\nonu
\A \A 
m_\theta^2 = m_\varphi^2 = m_\chi^2 = m_\nu^2 
= e^2 k^2 = -{{\xi e^2} \over {\kappa f}}, 
\nonu
\A \A 
m_{v_{a}} = m_{\omega} = 0,
\ea
which can produce {mass hierarchy} by the factor 
%
${e \over f}$.  
%
%
The local $U(1)$ gauge symmetry is not broken. 
The massless scalar $\omega$ is a NG boson for the degeneracy of the vacuum in $(A,\tilde B_{2})$-space, 
which is gauged away provided the gauge symmetry between the vector and the scalar multiplet is introduced. 

From these arguments we conclude that $N = 2$ SUSY QED is equivalent to $N = 2$ NLSUSY action, i.e. 
the matter sector (in asymptotic flat space) of $N = 2$ SGM 
produced by Big Decay (phase transition) of $N = 2$ NLSUSY GR (new space-time), 
possesses two different vacua, the type (a) and (b) in the $SO(3,1)$ isometry of (II) and (III). 
The resulting models describe 
two charged chiral fermions, two neutral chiral fermions, one massive vector, 
one charged massive scalar and one massless scalar: 
$(\psi_L{}^{cj} \sim (\tilde \chi_{DL}, \tilde \nu_{DL}), 
\lambda_L{}^j \sim \tilde \lambda_{DL}^j, v_a, 
\phi^c \sim \theta + i \omega, \phi^0 \sim \rho; j = 1, 2)$ for type (a) 
where ($\chi, \nu, \lambda$) are written by left-handed Dirac fields, 
and one charged Dirac fermion, one neutral (Dirac) fermion, a photon, 
one charged scalar 
and one neutral complex scalar (two neutral scalars): 
$(\psi_D{}^c \sim \chi + i \nu, \lambda_D{}^0 \sim \lambda^1 - i \lambda^2, v_a, 
\phi^c \sim \theta + i \varphi, \phi^{0c} \sim \rho + i \omega)$ for type (b), 
which are the composites of superons. 

As for cosmological significances of $N = 2$ SUSY QED in the SGM scenario, 
the vacuum of the cases (IIb) and (IIIb) produces the same interesting predictions 
as already pointed out in $N = 2$ pure SUSY QED in the SGM scenario \cite{ST4}, 
which may simply explain the observed mysterious (numerical) relations and give a new insight into the origin of mass 
\begin{center}
$((dark)\ energy\ density\ of\ the\ universe)_{obs} \sim (10^{-12})^4 \sim (m_\nu)_{obs}{}^4 
\sim {\Lambda \over G} \sim g_{sv}{}^2$, 
\end{center}
provided $-f \xi \sim O(1)$ and $\lambda^i$ is identified with neutrino. 
($\Lambda$, $G$ and $g_{sv}$ are the cosmological constant of NLSUSY GR (SGM) on {\it empty} new space-time for {\it everything}, 
the Newton gravitational constant and the superon-vacuum coupling constant via the supercurrent, 
respectively \cite{KS2,ST4}.) 
While the vacua of the cases (IIa) and (IIIa), 
equipped with automatic mixings of fermions in $d = 2$ so far,  
give new features characteristic of $N = 2$.  
They may be generic for $N > 2$ and deserve further investigations. 
%
%

The similar investigation in $d = 4$ is urgent and the extension to large $N$, especially to $N = 5$  
is important for {\it superon\ quintet\ hypothesis} in SGM scenario 
with ${N = \underline{10} = \underline{5}+\underline{5^{*}}}$ \cite{KS3} 
and to $N = 4$ is suggestive for analyzing the anomaly free nontrivial $d = 4$ field theory.     \par
Also NLSUSY GR in extra space-time dimensions is an interesting problem, 
which can describe all observed particles as elementary {\it \`{a} la} Kaluza-Klein. 

Our analysis shows that the vacua of the $N$-extended NLSUSY GR action in the SGM scenario possess rich structures 
promissing for the unified description of nature, where $N$-extended LSUSY theory appears 
as the vacuum field configurations of $N$-extended NLSUSY theory on Minkowski tangent space-time.

\vspace{15mm}

\noindent
One of the authors (K.S.) would like to express his sincere thanks to Julius Wess 
for his continuous warm hospitality and enlightening and encouraging discussions throughout these works, so long.  \\ 
Also K.S. and M.T. would like to thank N.S. Baaklini for his encouraging communications and interest in our works.

\newpage
\newcommand{\NP}[1]{{\it Nucl.\ Phys.\ }{\bf #1}}
\newcommand{\PL}[1]{{\it Phys.\ Lett.\ }{\bf #1}}
\newcommand{\CMP}[1]{{\it Commun.\ Math.\ Phys.\ }{\bf #1}}
\newcommand{\MPL}[1]{{\it Mod.\ Phys.\ Lett.\ }{\bf #1}}
\newcommand{\IJMP}[1]{{\it Int.\ J. Mod.\ Phys.\ }{\bf #1}}
\newcommand{\PR}[1]{{\it Phys.\ Rev.\ }{\bf #1}}
\newcommand{\PRL}[1]{{\it Phys.\ Rev.\ Lett.\ }{\bf #1}}
\newcommand{\PTP}[1]{{\it Prog.\ Theor.\ Phys.\ }{\bf #1}}
\newcommand{\PTPS}[1]{{\it Prog.\ Theor.\ Phys.\ Suppl.\ }{\bf #1}}
\newcommand{\AP}[1]{{\it Ann.\ Phys.\ }{\bf #1}}


\begin{thebibliography}{100}
\bibitem{WZ1}
J. Wess and B. Zumino, \PL{B49} (1974) 52. 

\bibitem{VA}
D.V. Volkov and V.P. Akulov, \PL{B46} (1973) 109. 

\bibitem{SS}
A. Salam and J. Strathdee, \PL{B49} (1974) 465. 

\bibitem{FI}
P. Fayet and J. Iliopoulos, \PL{B51} (1974) 461. 

\bibitem{O}
L. O'Raifeartaigh, \NP{B96} (1975) 331. 

\bibitem{KS1}
K. Shima, {\it Z. Phys.} {\bf C18} (1983) 25. 

\bibitem{CM}
S. Coleman and J. Mandula,  Phys. Rev.159 (1967) 1251.

\bibitem{HLS}
R. Haag, J. Lopszanski and M. Sohnius, Nucl. Phys. B88 (1975) 257.

\bibitem{WB} 
J. Wess and J. Bagger, {\it Supersymmetry and Supergravity} 
(Princeton University Press, Princeton, New Jersey, 1992). 

%

\bibitem{KS2}
K. Shima, \PL{B501} (2001) 237. 

\bibitem{KS3}
K. Shima, {\it European Phys. J.} {\bf C7} (1999) 341. 

\bibitem{ST3}
K. Shima and M. Tsuda, \PL{B507} (2001) 260. 

\bibitem{ST4}
K. Shima and M. Tsuda, {\it PoS HEP2005} (2006) 011; \\
K. Shima and M. Tsuda, \PL{B645} (2007) 455. 

\bibitem{STS}
K. Shima, M. Tsuda and M. Sawaguchi, 
{\it Int. J. Mod. Phys.} {\bf E13} (2004) 539. 

\bibitem{IK}
E.A. Ivanov and A.A. Kapustnikov, {\it J. Phys.\ }{\bf A11} (1978) 2375. 

\bibitem{R}
M. Ro\v{c}ek, \PRL{41} (1978) 451. 

\bibitem{UZ}
T. Uematsu and C.K. Zachos, \NP{B201} (1982) 250. 

\bibitem{STT1}
K. Shima, Y. Tanii and M. Tsuda, \PL{B525} (2002) 183. 

\bibitem{STT2}
K. Shima, Y. Tanii and M. Tsuda, \PL{B546} (2002) 162. 


\bibitem{ST2}
K. Shima and M. Tsuda, \MPL{A22} (2007) 1085. 

\bibitem{ST5}
K. Shima and M. Tsuda, arXiv:0706.0063 [hep-th] (2007). 





\end{thebibliography}
\end{document}